\documentclass{PoS}

\usepackage{amsmath,slashed,multirow,cite}

\newcommand{\gl}[1]{{\eqref{#1}}}
\newcommand{\eg}{{\it{e.g.}~}}
\newcommand{\ie}{{\it{i.e.}~}}
\newcommand{\cf}{{\it{cf.}~}}

\title{Results in Precision Multiboson+Jet Phenomenology}

\ShortTitle{Results in Precision Multiboson+Jet Phenomenology}

\author{Francisco~Campanario\\
  Institute for Theoretical Physics, KIT, 76128 Karlsruhe, Germany
}
\author{
  \speaker{Christoph Englert}\\
  Institut for Particle Physics Phenomenology, Department of
  Physics,\\Durham University, Durham, United Kingdom, and\\
  Institut f\"ur Theoretische Physik, Universit\"at Heidelberg, Germany\\
  E-mail: \email{christoph.englert@durham.ac.uk}}

\author{Michael~Rauch\\
  Institute for Theoretical Physics, KIT, 76128 Karlsruhe, Germany
}

\author{Michael~Spannowsky\\
  Institut for Particle Physics Phenomenology, Department of
  Physics,\\Durham University, Durham, United Kingdom
}

\author{Dieter~Zeppenfeld\\
  Institute for Theoretical Physics, KIT, 76128 Karlsruhe, Germany
}

\abstract{We review recent results in precision multiboson+jet
  phenomenology at the LHC. We discuss strategies how to compute these
  processes at NLO QCD and examine the impact of the perturbative
  corrections on the expected phenomenology, especially in the context
  of anomalous gauge boson couplings searches.}

\FullConference{10th International Symposium on Radiative Corrections (Applications of Quantum Field Theory to Phenomenology) - Radcor2011\\
  September 26-30, 2011\\
  Mamallapuram, India}

\begin{document}

\section{Introduction}
The production of multiple electroweak bosons provides important
channels to test LHC data against the well-established Standard Model
(SM). As the mechanism of electroweak symmetry breaking is currently
unknown, precise predictions of electroweak gauge boson production
rates are very important for quantifying deviations from the SM.
While the identification of electroweak bosons, especially in
their leptonic decay channels, is well established  experimentally,
the theoretical uncertainties we face at hadron colliders
are manifold.

Total rates and differential distributions suffer from severe theoretical
ambiguities
if they are limited to the leading order approximation in the perturbative
series' expansion. The uncertainties intrinsic to fixed order
calculations are conventionally assessed by investigating variations
of renormalization and factorization scales. The size of these
theoretical uncertainties turns out to be particularly large when we
deal with a LO color singlet final state configuration. This is
realized in pure electroweak production processes such as $pp\to VV$
or $pp\to VVV$ ($V=W^\pm,Z,\gamma$). In these channels the truncation
of the perturbative series at LO means that we force QCD to the odd
situation where no color gets transferred from the initial to the
final state. In other words, the next-to-leading order QCD corrections
to $pp\to VV,VVV$ \cite{vv,vvv,bozzi:waa} are large due to kinematically and
dynamically unsuppressed initial state parton radiation at large
center-of-mass energies. In fact, extra parton emission (\ie the real
emission contribution) is required for the cross sections' infrared
safety only in the soft and collinear limit. Hence, a reasonable
question to ask is whether we should include hard and resolvable jet
emission at all. Vetoing the additional jet activity, \ie
computing the exclusive hadronic cross section to next-to-leading
order, has been shown to quickly produce unreliable results
\cite{Campanario:2010hv,Campanario:2010xn,Campanario:2010hp,
  Campanario:2011ud} and should therefore be considered as a path which
should be avoided phenomenologically. Even worse from a theoretical
point of view, an appropriate choice of the jet veto scale can be
utilized to balance the scale dependencies of the (IR-regulated) real
emission contribution with the remaining parts of the NLO cross
section.  For \eg NLO QCD $W\gamma$ production processes
($\sqrt{s}=14~{\rm{TeV}}$) this is achieved by an experimentally
reasonable veto scale choice of $p_T^{\rm{veto}}\simeq 50~{\rm{GeV}}$.

However, many search strategies for new physics in
the context of multiboson production build upon specific final state
kinematical configurations and features of the scattering amplitude
which are special to the leading order approximation only. An example
is the search for anomalous trilinear couplings in
$W^\pm\gamma$ production \cite{Baur:1993ir} via the identification of the
so-called radiation zero \cite{Brown:1982xx}, which is a coherence
effect subject to distortion when additional parton radiation is taken
into account \cite{Diakonos:1992qc}.

Therefore, two serious limitations are seemingly avoided by imposing a
jet veto in anomalous couplings searches: (i) the cross section seems
to be perturbatively stable, and (ii), the radiation zero (and the
sensitivity to anomalous couplings in total) improves compared to the
inclusive NLO computation. However, this comes at the price of an uncertain
theoretical prediction which can not be trusted at the currently known
order of the perturbative series' expansion.

From this perspective the computation of multiboson+jet cross sections
is important for two reasons. On the one hand, adding NLO precision to
the one-jet--inclusive production provides an important piece of the
full NNLO multiboson production. On the other hand, theoretical
uncertainties of the vetoed cross section can be addressed with
sufficient precision. As a side benefit we can precisely analyze the
potential sensitivity to anomalous couplings which arises from recoils
against important initial state radiation.

\section{Inclusive searches for anomalous couplings}
We consider the most general $C$ and $P$ invariant extensions of the
electroweak gauge sector \cite{Hagiwara:1986vm}, which modifies the
trilinear $WW\gamma$ vertex
\begin{equation}
  \label{anovertex}
  \mathcal{L}_{{{WW}}\gamma} =
  -ie \big[ W_{\mu\nu}^\dagger W^\mu A^\nu- W_\mu^\dagger A_\nu W^{\mu\nu} 
  -\kappa_\gamma W_\mu^\dagger W_\nu F^{\mu\nu}
  +{\lambda_\gamma\over m_W^2} W_{\lambda\mu}^\dagger W^\mu_\nu
  F^{\nu\lambda}\big]\,
\end{equation}
and the trilinear $WWZ$ vertex
\begin{equation}
  \label{anovertexZ}
  \mathcal{L}_{{{WWZ}}} = 
  -ie \cot \theta_w \,\big[ g_1^Z\left( W_{\mu\nu}^\dagger W^\mu
    Z^\nu- 
    W_\mu^\dagger Z_\nu W^{\mu\nu}\right)
  -\kappa_Z W_\mu^\dagger W_\nu Z^{\mu\nu} +{\lambda_Z\over m_W^2}
  W_{\lambda\mu}^\dagger W^\mu_\nu Z^{\nu\lambda}\big]\,.
\end{equation}
%%%%%%%%%%%%%%%%%%%%%%%%%%%%%%%%%%%%%
\begin{figure}[!t]
  \begin{center}
%    \vspace{-0.3cm}
    \parbox{0.32\textwidth}{\includegraphics[scale=0.35]{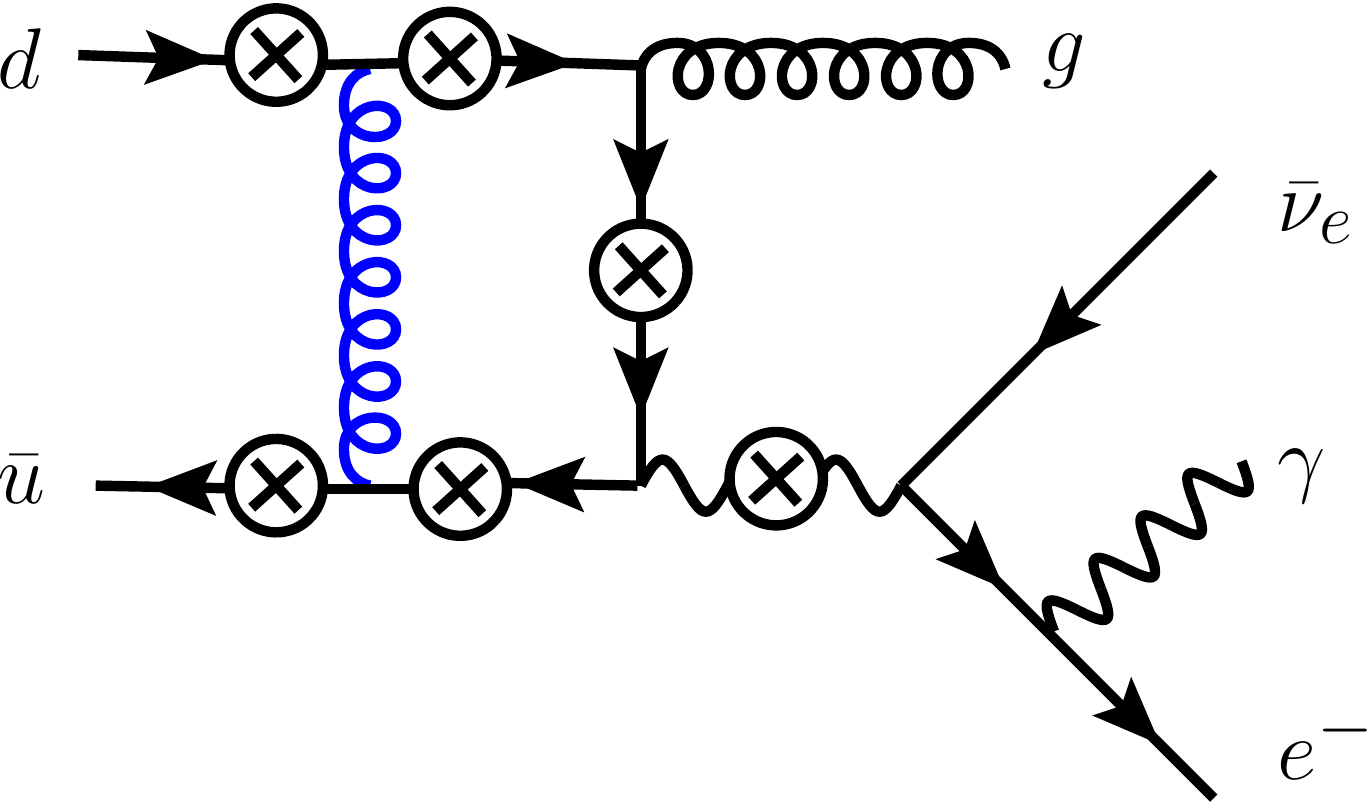}
    }
    \hspace{1.5cm}
    \parbox{0.32\textwidth}{\includegraphics[scale=0.35]{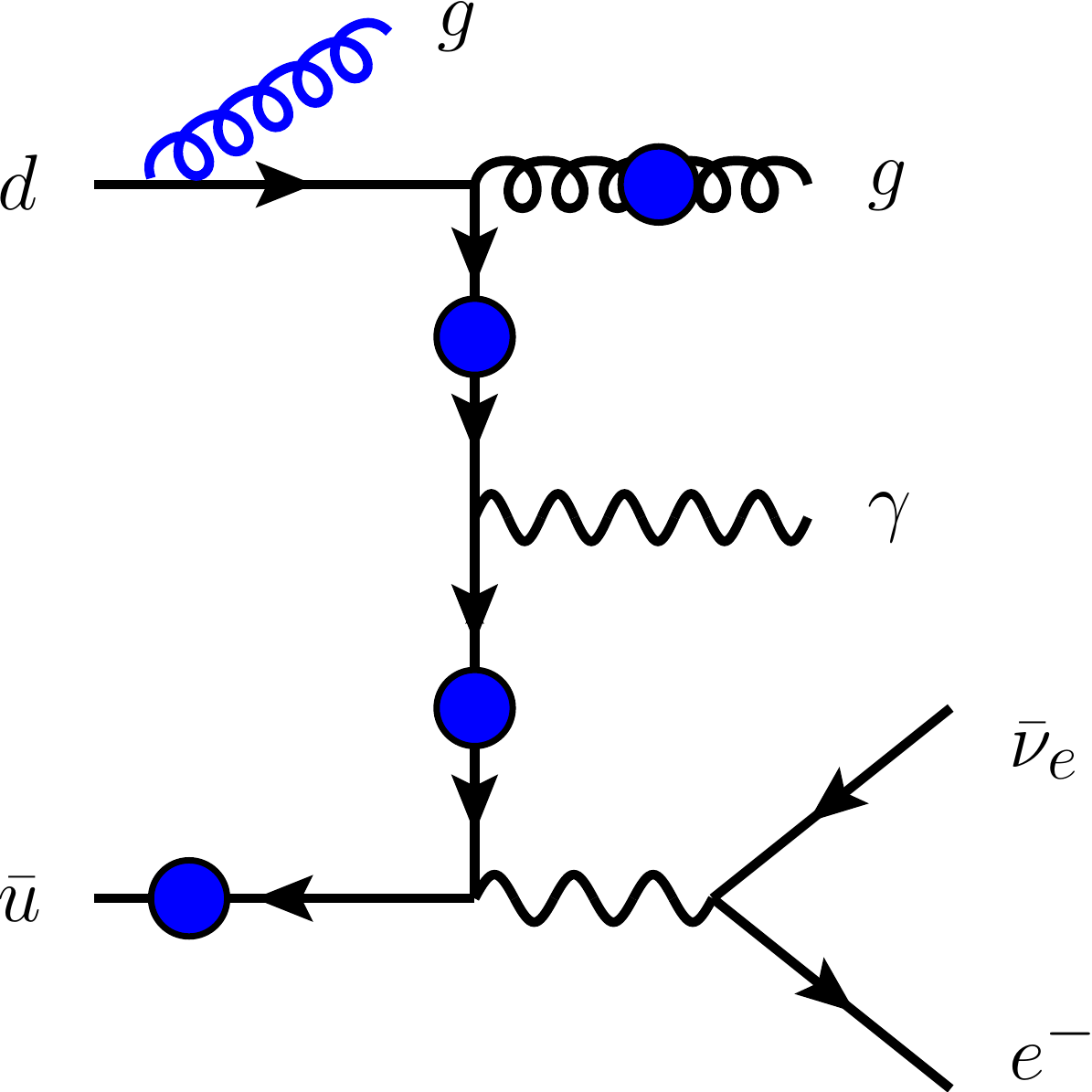}
    }
    \caption{\label{gl:feynwaj} Sample Feynman graphs contributing to
      $pp\to W\gamma$+jet production at NLO QCD. Left: Born and
      virtual contributions (the crosses mark possible photon
      couplings). Right: real emission contribution. The $pp\to
      WZ$+jet Feynman graphs follow by replacing the photon with the
      effective $(\gamma^\star,Z)$ decay current. The figures are
      taken from Ref.~\cite{Campanario:2010hv}.}
  \end{center}
%  \vspace{-0.5cm}
\end{figure}
%%%%%%%%%%%%%%%%%%%%%%%%%%%%%%%%%%%%%
Unitarity of cross sections at high energy scales requires the anomalous
parameters $(\kappa_\gamma,\kappa_Z,\lambda_\gamma,\lambda_Z, g_1^Z)$
to be understood as formfactors with a momentum dependence such that
for large momentum transfers we recover the SM from
Eqs.~\gl{anovertex} and \gl{anovertexZ}. Conventional choices are
dipole profiles~\cite{lepteva}, \eg
$\kappa_\gamma=1+\Delta\kappa_0/(1+m_{W\gamma}^2/\Lambda^2)$, where
$m_{W\gamma}$ denotes the mass of the invariant $W\gamma$ system and
$\Lambda$ relates to the cut-off scale underlying the effective theory of
Eq.~\gl{anovertex}.  The anomalous couplings modify a subset of
Feynman graphs which contribute to $pp\to \ell^\pm \slashed{p}_T
\gamma+{\rm{jet}}+X$ and $pp\to \ell^\pm \ell'^\pm
\ell'^\mp\slashed{p}_T \gamma +{\rm{jet}}+X$. The contributing Feynman
graphs at NLO QCD are indicated in Fig.~\ref{gl:feynwaj}.

We compute the NLO hadronic cross
section by straightforward application of the Catani-Seymour dipole
subtraction \cite{cs}. The loop contributions are evaluated using the
Passarino-Veltman scheme up to four-point functions \cite{pave} and
the Denner-Dittmaier reduction \cite{ddred} for five point integrals
and we perform various cross checks to validate our implementation
(for details see Refs.~\cite{Campanario:2010hv,
  Campanario:2010xn,Campanario:2010hp,Campanario:2011ud}).
%%%%%%%%%%%%%%%%%%
\begin{figure}[!t]
  \begin{center}
%    \vspace{-0.3cm}
    \parbox{0.55\textwidth}{\includegraphics[scale=0.9]{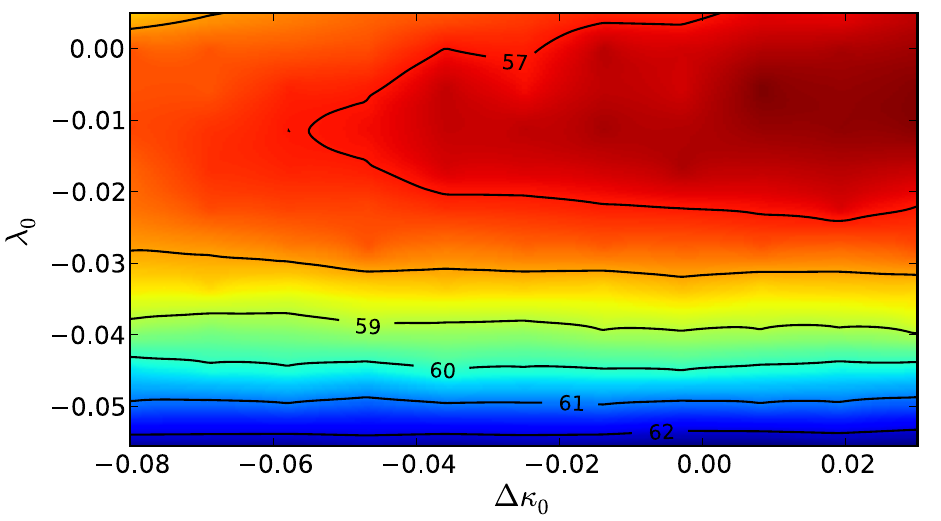}
    }
    \hspace{2mm}
    \parbox{0.4\textwidth}{\includegraphics[scale=0.9]{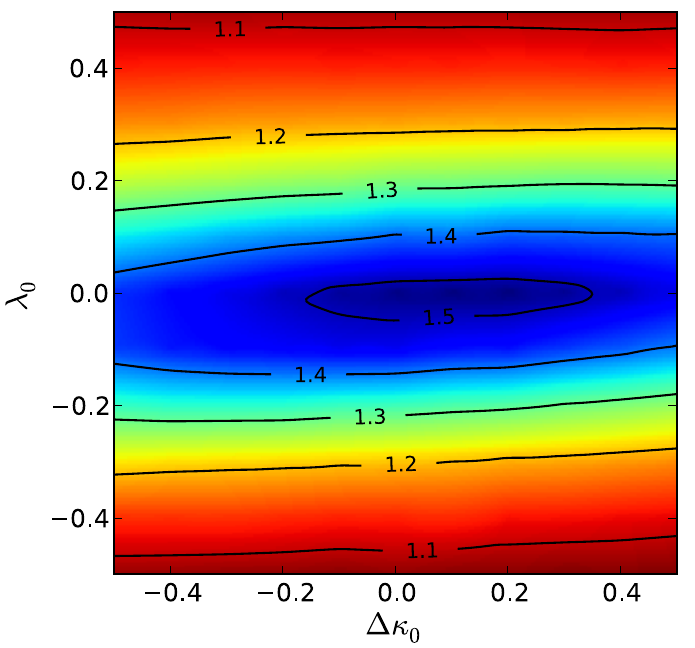}
    }
    \caption{\label{incl} Inclusive NLO QCD $pp \to
      \ell^-\slashed{p}_T\gamma+\hbox{jet}+X$ cross section (left, in fb) and
      K-factor contours (right)
      at the LHC after applying typical search cuts, as a function of
      anomalous coupling parameters
      \cite{Campanario:2010hv,dobbs,lepteva}. The cutoff scale in the
      dipole form factor is chosen as $\Lambda=2~{\rm{TeV}}$. Figures
      taken from Ref.~\cite{Campanario:2010hv}.}
  \end{center}
%  \vspace{-1cm}
\end{figure}
%%%%%%%%%%%%%%%%%%
In Fig.~\ref{incl} we show the total inclusive $pp \to
\ell^-\bar\nu_\ell\gamma$+jet cross section for anomalous
production. The chosen cuts are adopted from typical searches for
anomalous couplings which seek to project out the phase space regions
with large momentum transfers in the $WW\gamma$ ($WWZ$) couplings,
thus enhancing the sensitivity to anomalous trilinear gauge
couplings. In particular, this amounts to hard photons
($p_T^\gamma\gtrsim 100$ GeV) or $Z$s recoiling against the $W$s (in
the following we focus on the $WW\gamma$ vertex
\cite{Campanario:2010hv}, for further details on $WWZ$ see
Ref.~\cite{Campanario:2010xn}).

%%%%%%%%%%%%%%%%%%%%%%%%%%%%%%%%%%%%%%%%%%%%
\begin{figure}[!b]
  \begin{center}
    \includegraphics[height=5.5cm]{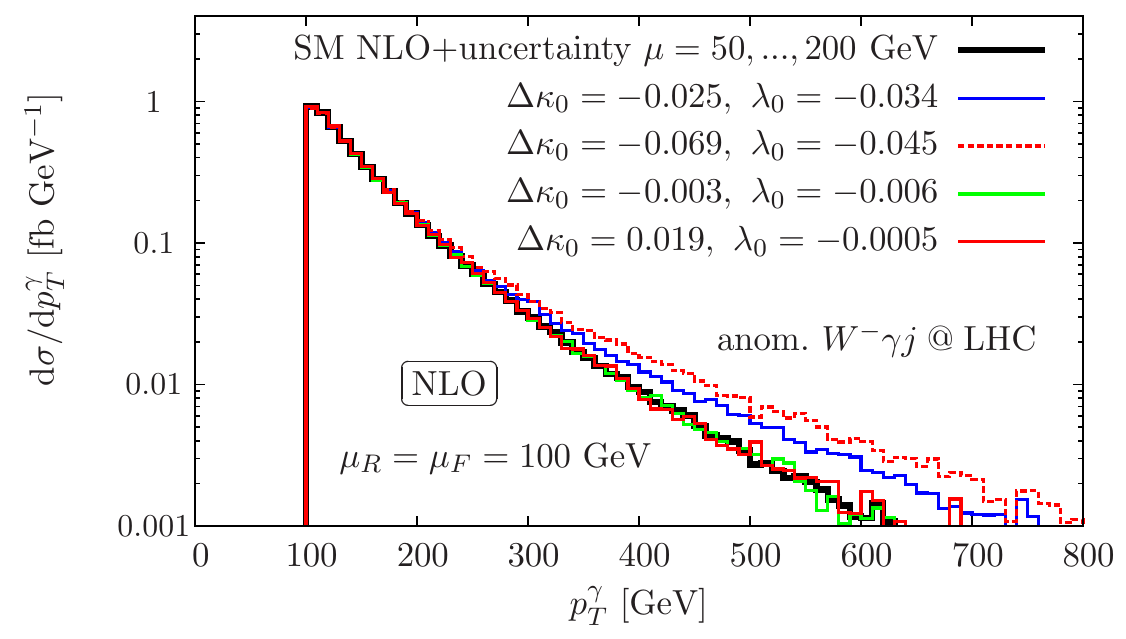}
    \caption{\label{diffa} Photon transverse momentum distribution and
      uncertainty band for SM and anomalous production at NLO QCD from
      Ref.~\cite{Campanario:2010hv}. Dynamical scale choices have no
      sizable impact.}
  \end{center}
%  \vspace{-0.7cm}
\end{figure}
%%%%%%%%%%%%%%%%%%%%%%%%%%%%%%%%%%%%%%%%%%%%

The QCD corrections turn out to be most sizable for anomalous
couplings choices close to the SM ($\lambda_i=0$, $\kappa_i=1$. This is a
consequence of the differential $K$ factor being large around the
$p_T^\gamma$ threshold while less significant in the distribution's
tail. Therefore, large anomalous couplings choices, for which the
cross section is significantly enhanced in the $p_T$ tails, receive a
less sizable relative correction compared to SM-like choices. The
photon's transverse momentum distribution, Fig.~\ref{diffa}, is
typically employed to establish exclusion bounds on anomalous
couplings \cite{dobbs}.  Even though the (rather) inclusive NLO cross
sections' scale dependences of $\sim 10\%$ turn out to be comparable to
the relative difference induced by anomalous couplings (\cf
Fig.~\ref{incl}), the distributions reveal a significant dependence on
the anomalous parameters, especially on $\lambda_i$ in
Eqs.~\gl{anovertex},~\gl{anovertexZ}.
We perform a binned log-likelihood hypothesis test (taking into
account the shape uncertainty that arises from QCD scale uncertainty
as a nuisance parameter), \cf Ref.~\cite{Read:2002hq}, to estimate to
what extent inclusive searches can help to improve the
currently pursued analyses strategies\footnote{Note that for very hard
  photons $p_T^\gamma\geq 100$ GeV there is only minimal pollution of
  misidentified jets and our analysis is essentially background-free
  \cite{dc1}.}. We find that parameter choices at the edge of the
currently allowed LEP bounds on anomalous $WW\gamma$ interactions can
be excluded at the $3\sigma$ level for ${\cal{L}}\simeq
25~{\rm{fb}}^{-1}$ running the LHC at $14$ TeV. Note that this result
is partly due to the small uncertainty band which follows from the
higher NLO precision of the $p_T$ distributions.

All results are obtained with an updated version of the {\sc{Vbfnlo}}
package \cite{Arnold:2008rz}, which is publicly available.

\section{$W\gamma\gamma$+jet production at NLO: Toward anomalous
  quartic couplings}
Among the triple vector boson production channels, $W\gamma\gamma$
has turned out to be of particular interest. It is the channel with
the largest $K$ factor ($\sim 4$) for the integrated cross section, due to the radiation
zero, and it is also a very promising channel to measure anomalous
quartic couplings~\cite{quartic}. The remaining scale uncertainties at
NLO QCD~\cite{bozzi:waa} are due to unbalanced gluon-induced real
radiation contributions appearing first at NLO, which are computed at
LO, e.g., $gq\to$ $W\gamma\gamma~q$.  The observation of the radiation
zero is also obscured, similar to $W\gamma$ production, by additional
real QCD radiation, $W\gamma\gamma$+jet, as part of the NLO
calculation. In Ref.~\cite{bozzi:waa} it was shown that an additional
jet veto-cut might help in the detection of the radiation zero making
visible the dip and also reducing the scale uncertainties for the
relevant distributions. However, this procedure raises the question of
the reliability of the predictions due to the aforementioned problem
with the exclusive vetoed samples. To realistically assess the
uncertainties, also concerning anomalous coupling searches, and as an
important step towards a NNLO QCD calculation of $W\gamma\gamma$, we
have calculated $W\gamma\gamma$+jet production at NLO QCD. This is the
first calculation falling in the category of $VVV$+jet production and
includes the evaluation of the complex hexagon virtual amplitudes,
which poses a challenge not only at the level of the analytical
calculation, but also on the level of required CPU time computing the
full $2\to 4$ NLO matrix element.

For the virtual contributions we use the routines computed in
Ref.~\cite{Campanario:2011cs} which employ \textsc{FeynCalc}
\cite{Mertig:1990an} and {\sc FeynArts} \cite{Hahn:2000kx} in an
in-house framework. At the numerical evaluation level, we split the
virtual contributions into fermionic loops~(Virtual-fermionbox) and
bosonic contributions with one~(Virtual-box), two~(Virtual-pentagons)
and three~(Virtual-hexagons) electroweak vector bosons attached to the
quark line.  This procedure allows us to drastically reduce the time
spent in evaluating the part containing hexagon diagrams as explained
in Refs.~\cite{Campanario:2011cs,Campanario:2011ud}. The numerical
stability of the hexagons' diagram evaluation is discussed in detail
in Ref.~\cite{Campanario:2011cs}.

\subsection{Results}
\begin{figure}[!t]
  \begin{center}
%    \vspace{-0.3cm}
    \includegraphics[width=\textwidth]{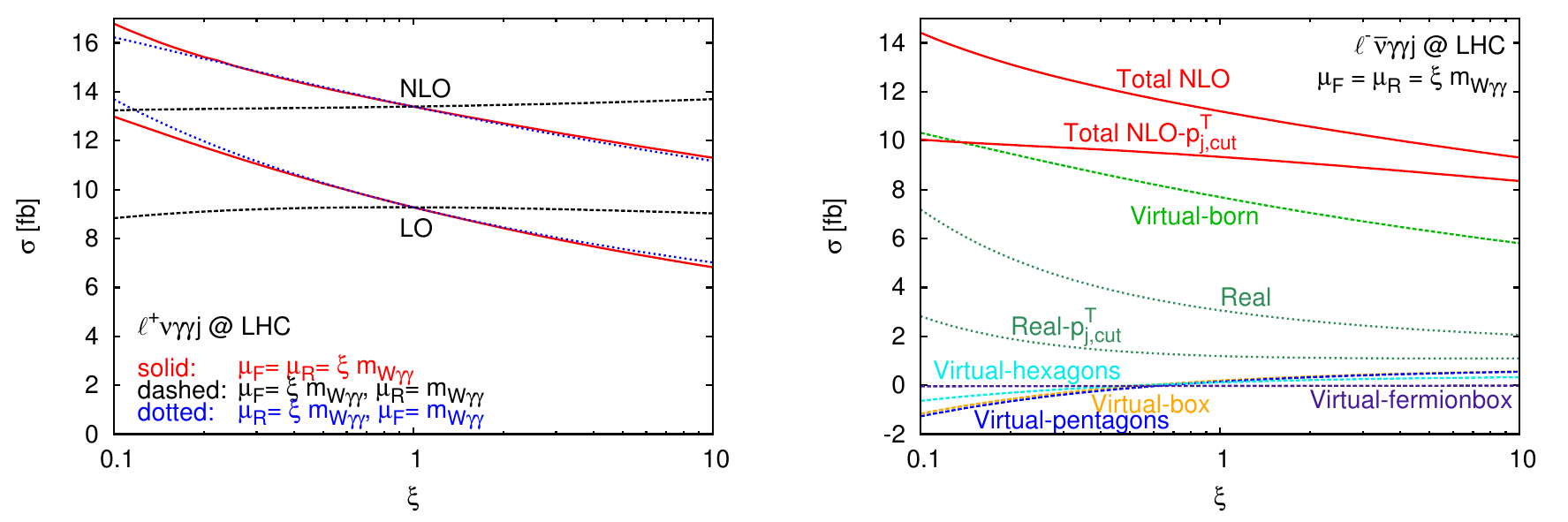}
    \caption{\label{fig:waaj:scale} Scale variation of the $\ell^\pm
      \nu \gamma\gamma$+jet production cross sections at the LHC
      ($\ell$ = $e$, $\mu$) taken from
      Ref.~\cite{Campanario:2011ud}. The cuts are described in the
      text and we choose $\mu_R=\mu_F=m_{W\gamma\gamma}$ as central
      dynamical reference scale. The right panel shows the individual
      contributions to the NLO cross section according to our
      classification of topologies. We also show results where we have
      applied a veto on events with two identified jets having both a
      transverse momentum larger than 50 GeV.}
  \end{center}
%  \vspace{-0.5cm}
\end{figure}

\begin{figure}[!t]
  \begin{center}
    \includegraphics[height=8cm]{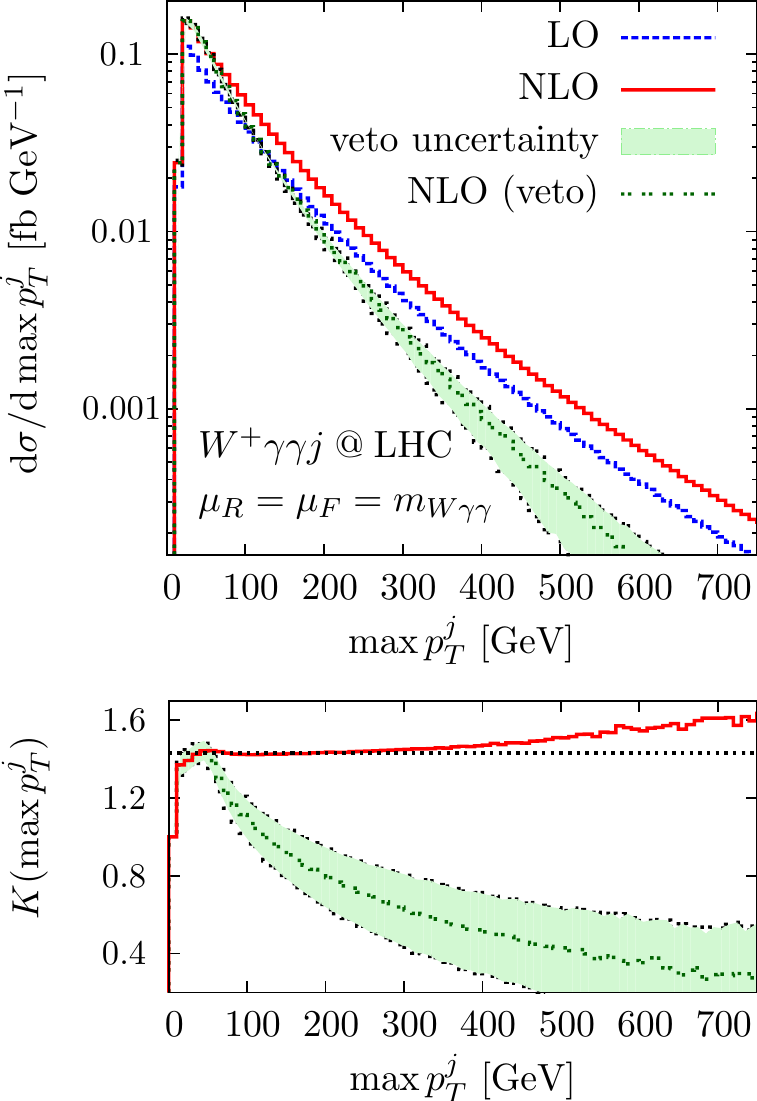}
    \hspace{2cm}
    \includegraphics[height=8cm]{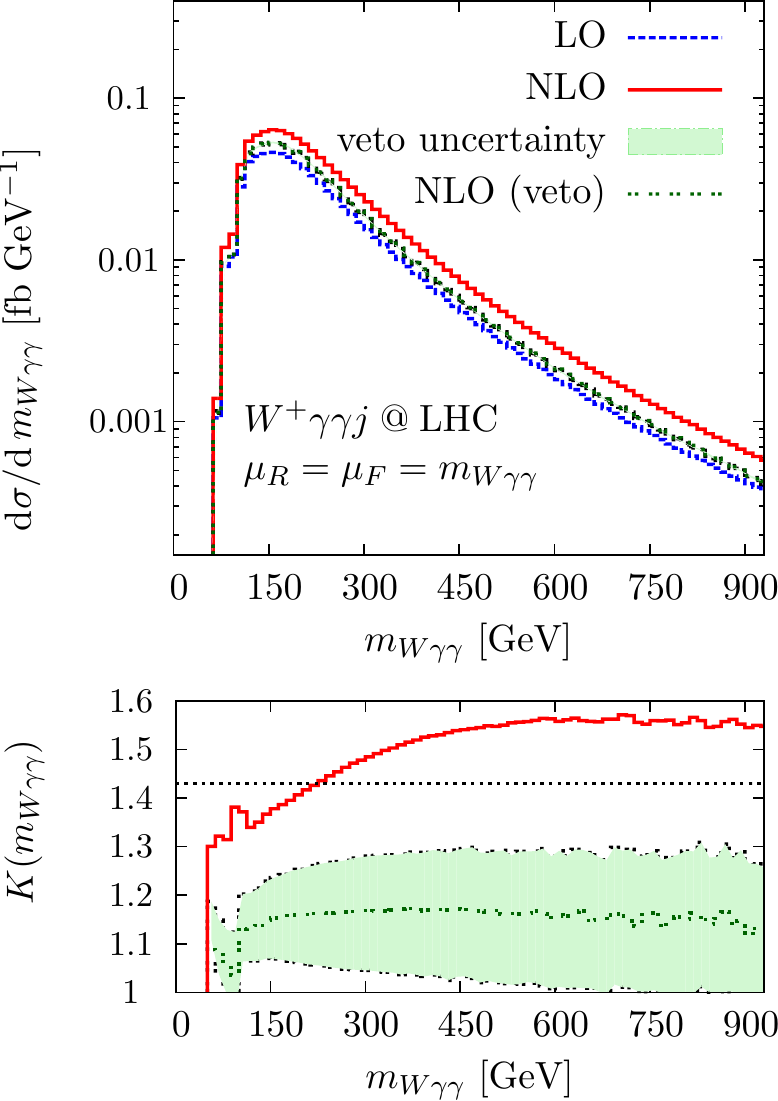}
    \caption{\label{fig:waaj:diff} Differential $\max p_T^j$ and
      $m_{W\gamma\gamma}$ distribution for inclusive and exclusive
      $W^-\gamma\gamma$+jet production. Figures taken from
      Ref.~\cite{Campanario:2011ud}.}
  \end{center}
\end{figure}

For the cross sections and distributions of Tab.~\ref{tab:xsecs} and
Figs.~\ref{fig:waaj:scale} and \ref{fig:waaj:diff} we use the CT10 parton
distribution set \cite{Lai:2010vv} with $\alpha_s(m_Z)=0.118$ at NLO,
and the CTEQ6L1 set~\cite{Pumplin:2002vw} with $\alpha_s(m_Z)=0.130$
at LO. Further details on the parameter choices can be found in
Ref.~\cite{Campanario:2011ud}.  Again, we consider $W^\pm$ decays to
the first two lepton generations, \ie the decays $W\rightarrow
e\nu_e,\mu\nu_\mu$ and these
contributions have been summed in Fig.~\ref{fig:waaj:diff}.

We choose inclusive cuts to study the impact of the inclusive QCD
corrections in a general setting: $p_T^{j}\geq 20~\rm{GeV}$,
$p_{T}^\ell \geq 20 ~\rm{GeV}$ (10~GeV at the Tevatron), $p_{T}^\gamma
\geq 20 ~\rm{GeV}$ (10~GeV at the Tevatron),
$|\eta_\ell|,|\eta_\gamma|\leq 2.5$, and an azimuthal
angle-pseudorapidity plane separation $R_{\ell\gamma} = (\Delta
\phi_{\ell\gamma}^2 + \Delta\eta_{\ell\gamma}^2)^{1/2}\geq 0.4$.  For
the separation of the charged lepton from observable jets, we choose
$R_{\ell j}\geq 0.4$ and we require $R_{\gamma\gamma}\geq 0.4$ for the
diphoton separation.  Besides the photon-parton isolation criterion
according to Ref.~\cite{Frixione:1998jh} with the separation parameter
$\delta_0=0.7$, we also require a separation between photons and
identified jets of $R_{\gamma j}\geq 0.7$.

%%%%%%%%%%%%%%%%%%%%%%%%%%%%%%%%%%%%%
%%%%%%%%%%%%%%%%%%%%%%%%%%%%%%%%%%%%%
%%%%%%%%%%%%%%%%%%%%%%%%%%%%%%%%%%%%%
%%%%%%%%%%%%%%%%%%%%%%%%%%%%%%%%%%%%%
%%%%%%%%%%%%%%%%%%%%%%%%%%%%%%%%%%%%%
\begin{table}[!t]
  \begin{center}
    \begin{tabular}{c c c c c}
      \hline
      \hline
      & $\sigma^{\rm{LO}}$ [fb] & $\sigma^{\rm{NLO}}$ [fb] & $K=\sigma^{\rm{NLO}}/\sigma^{\rm{LO}}$ & \\ 
      \hline
      \hline
      $W^\pm\gamma\gamma$+jet & 1.191 &    1.754  & 1.47 & Tevatron \\
      \hline
      $W^+\gamma\gamma$+jet &  4.640  &   6.634  & 1.43 &  \multirow{2}{*}{LHC} \\
      $W^-\gamma\gamma$+jet &   3.803 &  5.644 &  1.48 & \\
    \end{tabular}
    \caption{\label{tab:xsecs} Total LO and NLO cross sections and $K$
      factors for
      $\stackrel{}{p}\stackrel{\text{\tiny(}-\text{\tiny)}}{p}\to
      e^-\bar\nu_e\gamma\gamma$+jet+$X$ and
      $\stackrel{}{p}\stackrel{\text{\tiny(}-\text{\tiny)}}{p}\to
      e^+\nu_e\gamma\gamma$+jet$+X$ at the Tevatron and at the
      LHC. Results taken from Ref.~\cite{Campanario:2011ud}.}
  \end{center}
%  \vspace{-0.5cm}
\end{table}
%%%%%%%%%%%%%%%%%%%%%%%%%%%%%%%%%%%%%

We compute total $K$ factors of 1.43 (1.48) for $W^+\gamma\gamma$+jet
($W^-\gamma\gamma$+jet) production at the LHC, values which are quite
typical for multiboson+jet production as found in
Refs.~\cite{Campanario:2010hv, Campanario:2010xn,dibosjet}. This
moderate $K$ factor (as compared to corrections of $\sim
400\%$ for $W\gamma\gamma$ production) indicates, as expected, that the
$W\gamma\gamma$+jet production channel is not affected by radiation
zero cancellations, since the
bulk of the $W\gamma\gamma$+jet cross section is due to gluon induced
processes which do not sport a radiation zero.
 
The scale dependence of the $W^+\gamma\gamma j$ and $W^-\gamma\gamma
j$ production cross sections turns out to be modest: when comparing
$\mu_R=\mu_F=\xi m_{W\gamma\gamma}$ for $\xi=0.5$ and $\xi=2$, we find
differences of $10.8\%$ ($12.0\%$), respectively.

The phase space dependence of the QCD corrections is non-trivial and
sizable (we again choose $\mu_R=\mu_F=m_{W\gamma\gamma}$). Vetoed
real-emission distributions are plagued with large uncertainties --- a
characteristic trait well-known from $VV$+jet phenomenology
\cite{dibosjet,Campanario:2010hp}. Additional parton emission modifies
the transverse momentum and invariant mass spectra among other
distributions such as minimum separations, etc. The leading jet becomes
slightly harder at NLO as can be inferred from the differential $K$
factor in the bottom panel of Fig.~\ref{fig:waaj:diff}. When comparing
precisely measured distributions in this channel against LO Monte
Carlo predictions, the not-included QCD corrections could be
misinterpreted for anomalous electroweak trilinear or quartic
couplings \cite{Campanario:2010hv, Campanario:2010xn, Baur:1993ir}
arising from new interactions beyond the SM.

\end{document}